# Experimental Study of the Nematic Transition in Granular Spherocylinder Packings under Tapping

Haitao Yu,[1] Zhikun Zeng,[1] Ye Yuan,[1] Shuyang Zhang,[1] Chengjie Xia,[4,*] and Yujie Wang[1,2,3,†]

[1]*School of Physics and Astronomy, Shanghai Jiao Tong University, 800 Dong Chuan Road, Shanghai 200240, China*
[2]*State Key Laboratory of Geohazard Prevention and Geoenvironment Protection, Chengdu University of Technology, Chengdu 610059, China*
[3]*Department of Physics, College of Mathematics and Physics, Chengdu University of Technology, Chengdu 610059, China*
[4]*School of Physics and Electronic Science, East China Normal University, Shanghai 200241, China*

Using x-ray tomography, we experimentally investigate the nematic transition in granular spherocylinder packings induced by tapping. Upon the validation of the Edwards ensemble framework in spherocylinders, we introduce an empirical free energy that accounts for the influence of gravity and the mechanical stability requirements specific to granular systems. This free energy can predict not only the correct phase transition behavior of the system from a disordered state to a nematic phase, but also a phase coexistence range and nucleation energy barriers that agree with experimental observations.

The packing of hard particles is ubiquitous in nature and industrial processes [1,2], which can be traced back to Kepler's research on ball packing in 1611 [3]. Since then, the structure and phase behavior of sphere packings have been extensively studied. In recent decades, growing studies on packings of non-spherical particles have revealed appealing richness in shape-dependent packing properties and phase behaviors [4-10]. A classic example is the isotropic-nematic transition observed in hard-rod systems owing to the competition between orientational and translational entropy, as explained by Onsager's excluded volume theory [11]. Subsequent numerical and theoretical studies have extensively investigated the equilibrium phase boundaries for hard-rod systems [12-15]. It is worth noting that most of these works have primarily focused on thermal systems, such as colloidal rods [16-19]. As for the out-of-equilibrium granular systems, similar behaviors as their thermal counterparts have been observed. For example, long granular rods, under external agitation, tend to align and lead to an abrupt densification characterized by nematic-like and smectic-like orderings [20-29]. However, due to the intrinsic athermal nature of granular materials, it remains unclear whether their phase behaviors and associated mechanisms can be directly mapped onto thermal systems [30]. Previously, Galanis et al. [31] utilized a generalized thermal-equilibrium free-energy minimization approach to elucidate the phase separation transition in a highly-agitated two-dimensional mixture of granular rods and spheres. However, there is no straightforward application of the thermal-equilibrium statistical mechanics in the case of static granular packings. Instead, an analogous statistical mechanical framework has been proposed by Edwards and coworkers [32], whose validity has been successfully tested in spherical granular packings recently [33,34]. Using this framework, Ding et al. [35] explained a novel cubatic structural transformation in packings of granular cylinders with an aspect ratio close to one. Nevertheless, more research is needed to investigate the potential extension of this framework to other shapes and explore fundamental differences from thermal systems.

In this Letter, we employ x-ray tomography to investigate the compaction process of granular spherocylinder packings under tapping. Our results reveal a clear first-order nematic phase transition in the system. To provide an understanding of this observed transition, we first verify the applicability of the Edwards ensemble to spherocylinders. Subsequently, we quantify the transition by monitoring particle-scale structural transformation during the transition, which allows us to obtain the excluded volume and orientational entropy associated with each spherocylinder. Based on the Edwards framework, we construct a phenomenological mean-field free energy incorporating new terms that account for the influence of gravity and the mechanical stability requirements specific to granular systems, which are absent in thermal systems. Based on this modified free energy, we can accurately predict a first-order phase transition with a coexistence range that aligns with experimental observations. This free energy also predicts nucleation energy barriers consistent with those obtained from the distribution of the nucleating clusters based on the classical nucleation theory.

The samples used in this study are 3D-printed (ProJet MJP 2500 Plus) plastic spherocylinders, which consist of a cylindrical body with hemispherical caps at both ends. The diameter of the spherocylinders is $D = 4$ mm, the length of the cylindrical part is $L_{cylinder} = 24$ mm, and thus the aspect ratio $\alpha = (L_{cylinder} + D)/D = 7$. The packings are prepared in a cylindrical plastic container with a diameter of 180 cm and a height of about 100 cm. To minimize boundary effects, we glue on the inner surface of the container with segments of spherocylinders with random orientations and lengths ranging from $1D$ to $6D$. Each packing consists of about 5300 spherocylinders. When preparing the packing, we insert a thinner cylindrical tube into the container and gently pour spherocylinders into the tube. We then slowly withdraw the tube vertically, allowing the spherocylinders to fill and settle in the container gently. Following this procedure, we can obtain a reproducible loose packing. Other different packing densities can be realized by tapping the system by a mechanical shaker with tap intensities $\Gamma = 2g \sim 16g$, where $g$ is the gravitational acceleration constant. Each tapping cycle consists of a 300 ms pulse followed by a 1.5 s interval. The system is tapped for 1000~100,000 times, depending on $\Gamma$. The evolution of the packing structures under tapping is obtained by a medical CT scanner (UEG Medical Group Ltd., 0.2 mm spatial resolution). Following similar image processing procedures as previous studies [36,37], the centroid and orientation of each spherocylinder can be determined with uncertainties less than $0.01D$ and 0.1 degrees, respectively. In the subsequent analysis, we only include particles located at least $7D$ distance away from the container boundary.

To understand the structural changes and the associated phase transition process of the spherocylinder packing during tapping, we first examine the evolution of volume fraction $\phi = \langle v_p \rangle / \langle v \rangle$ under different $\Gamma$ as a function of tapping number $t$, where $v_p$ and $v$ are the respective volumes for the particles and their associated Voronoi cells [see Fig. 1(b)]. To characterize the orientational order of the packing, we employ the nematic order parameter $s = \langle 3\cos^2 \theta - 1 \rangle / 2$ from liquid crystal theory, where $\theta$ is the included angle between the particle orientation and the direction of gravity. Figures 1(b) and (c) show that all packings are initially in disordered states with $\phi = 47.6\% \pm 0.4\%$, defined as $s < -0.2$, in which most of the spherocylinders are lying horizontally in random orientations [upper of Fig. 1(a)]. As tapping is applied at different intensities, the system evolves differently. For $\Gamma \leq 6g$, the system experiences gradual compaction during tapping but remains in the disordered state throughout the experiment duration. For $\Gamma \geq 7g$, the system can reach a new stable state where $s \approx 0.5$, indicating the emergence of nematic ordering in the system [lower of Fig. 1(b)]. The associated volume fraction $\phi$ at the stable states decreases with the increase of tapping intensity. It is noteworthy that there is a sudden increase in both $\phi$ and *s* during the nematic phase transition, a characteristic feature of a typical first-order phase transition. For $\Gamma \approx 8g$, both the volume fraction and the order parameter abruptly increase with the further increase in tapping number, which demonstrates the appearance of smectic ordering in the system.

In order to gain a quantitative understanding of the observed phenomena, we treat the tapped granular spherocylinder systems using the Edwards ensemble framework, similar to the case of granular sphere packings [33]. Specifically, we calculate the Voronoi cell volume variance var(*v*) at different volume fractions, as shown in Fig. 2(a), where $v_p$ is set as unity for simplicity. We then fit this data using a quadratic polynomial [solid curve in Fig. 2(a)] to obtain an analytical expression, used to calculate the Edwards compactivity, which act as an effective temperature for granular packings. Note that we only use var(*v*) of the disordered branch [solid symbols in Fig. 2(a)], which exhibits a smooth continuation of the system's low $\phi$ behavior, since no clear one-to-one relationship between var(*v*) and $\phi$ can be identified when $\phi$ is about 0.52~0.6, where the coexistence of distinctive phases occurs. The compactivity of any packing at a specific $\phi$ is determined by the fluctuation method [33]:

$$\frac{1}{\chi(\phi)} = \int_{\phi_{rlp}}^{\phi} \frac{d\psi}{\psi^2 \, \text{var}(v)} . \qquad (1)$$

where $\phi_{rlp} \approx 0.46$ is the packing fraction of the random loose packing (RLP) state with an infinite $\chi$. Similar to spherical particles, we also adopt an alternative histogram overlapping method to calculate the compactivity $\chi$ [33], which can also verify the equal-probability assumption of the Edwards ensemble. According to this method, if the density of states sampled at different compactivity is identical for a same granular packing system, the logarithm of the ratio between the volume distribution $P(v)$ for packings at different $\chi$ should linearly depend on $v$. This phenomenon is indeed observed in our experiment, as shown in the inset of Fig. 2(b). Notably, the values of $\chi$ calculated using the histogram overlapping method agree well with those obtained using the fluctuation method, as shown in Fig. 2(b), providing further support to the validity of Edwards ensemble in non-spherical particle systems.

Once the validity of the Edwards ensemble framework is verified, we can proceed to obtain the free energy of the system to gain deeper insights into the nematic phase transition. In our spherocylindrical system, the single-particle free energy $f$ can be tentatively expressed to first order as [38]:

$$f(\chi) = W - \chi S_\theta, \qquad (2)$$

where $W$ is the packing volume, analogous to energy in a conventional thermal system, and $S_\theta$ is the orientational entropy. Here we assume that the nematic phase transition is driven primarily by the variation in orientational entropy, given that the change in translational entropies across the transition is usually small and can be neglected [35].

To develop a model encompassing both the packing volume and orientational entropy, we need to first obtain the orientational distribution of all spherocylinders in the system. For simplicity, we employ a mean-field approximation assuming that each spherocylinder independently satisfies the same orientational distribution function (ODF) as follows: $P(\theta) = k \exp\left(\sum_{i=2}^{n} \lambda_i P_i(\cos\theta)\right)$, where $\lambda_i$ is the Lagrangian multiplier for the $i$th-order Legendre polynomial $P_i$, and $i$ takes even values from 2 to $n$. We note that $\langle P_2 \rangle = \left\langle \frac{1}{2}(3\cos^2\theta - 1) \right\rangle$ is nothing but the order parameter $s$ along the gravity direction. Through the utilization of the maximum-entropy method, we can approximate the full ODF by employing a limited set of parameters [39]. Specifically, when we consider solely the first two terms, the ODF takes the following form of $P(\theta) = k(\lambda_2, \lambda_4)\exp(\lambda_2 P_2(\cos\theta) + \lambda_4 P_4(\cos\theta))$, in which each pair $(\lambda_2, \lambda_2)$ defines a complete set of $\langle P_i \rangle$ momenta and thus a given shape of the ODF. Furthermore, an empirical formula $\lambda_4 = f(\lambda_2)$ between $\lambda_2$ and $\lambda_4$ can be established by fitting the experimental data, allowing us to write the ODF as:

$$P(\theta) = k(\lambda_2)\exp(\lambda_2 P_2(\cos\theta) + f(\lambda_2) P_4(\cos\theta)). \tag{3}$$

The calculated ODF reproduce rather satisfactorily the experimental orientational distribution, as shown in Fig. 3(a).

Once the ODF is determined, we can calculate the excluded volumes between two spherocylinders when they are in contact at an angle $\gamma$, using Onsager's excluded volume theory:

$$V_{excl}(\gamma) = (4\pi/3)D^3 + 2\pi D^2 L + 2DL^2 \sin\gamma. \tag{4}$$

Integration over the ODF yields the average excluded volume of the system, as shown in Fig. 3(b):

$$\langle V_{excl} \rangle = \int P(\theta_1) P(\theta_2) V_{excl}(\gamma_{12}) d\Omega_1(\theta_1, \varphi_1) d\Omega_2(\theta_2, \varphi_2), \tag{5}$$

where $\gamma_{12}$ is the contacting angle between two particles, and $d\Omega$ denotes the integration over the full solid angle. In this calculation, we assume that the orientations of the spherocylinders are uniformly distributed in the other spherical coordinate $\varphi$. By adopting a simple random contact model [40], $W(\lambda_2)$ can be obtained from the excluded volume as $W = \frac{\langle V_{excl} \rangle}{2z}$, where $z$ is the average contact number (see Supplemental Materials [41] for more details). Furthermore, the orientational entropy can be calculated as [upper red curve in Fig. 3(b)]:

$$S_\theta = -\iint P(\theta) \log[4\pi P(\theta)] \sin\theta d\theta d\varphi, \tag{6}$$

where we have subtracted the entropy of the system when particles are uniformly distributed in orientation.

It is important to note that for granular systems, the free energy $f$ needs modification compared to its simplified thermal form of Eq. (2) due to the influence of gravity and mechanical stability constraints. Unlike thermal systems, granular systems with finite friction tend to have the spherocylinders lying horizontally over each other, as contacts orientated along other directions are less likely to maintain mechanical stability under gravity. This highlights a fundamental difference between the Edwards ensemble and the thermal ensemble, as the requirement of the mechanical stability of the Edwards ensemble introduces varying statistical weights for different contact configurations. To account for the preference for contacts to be orientated along the gravity direction, we modify the free energy $f$ by introducing a new term [lower red curve in Fig. 3(b)]:

$$S_z = S_{z0} \left\langle \left( \frac{\mathbf{u}_1 \times \mathbf{u}_2}{|\mathbf{u}_1 \times \mathbf{u}_2|} \right)_z^2 \right\rangle, \tag{7}$$

where the contact direction is calculated by the cross product of the orientations of the two contacting particles $\mathbf{u}_1 \times \mathbf{u}_2$, $S_{z0} = 2$ is an empirical fitting parameter and $z$ denotes the component along the vertical direction. Additionally, it is observed that when nematic order emerges in the system, the order tends to align with the gravity direction, indicating a coupling of the order parameter with the gravity field $hs$, where $h = 0.27$ is another fitting parameter characterizing the coupling strength. The modified free energy now becomes:

$$f(\chi, s) = W(s) - \chi \left( S_\theta(s) + S_z(s) \right) + hs. \tag{8}$$

It turns out the phase transition behavior of this empirical free energy $f$, as the system undergoes the transition from the disordered phase to the nematic phase, is in quantitative agreement with the experimental results [Fig. 3(d)]. Specifically, as shown in Fig. 3(c), when $\phi$ is small, there exists a single minimum at $S \approx -0.5$ in the free energy, corresponding to the disordered phase where most particles lie close to the horizontal plane. Around $\phi \approx 0.52$, two minima appear in the free energy, corresponding to the emergence of the ordered nematic phase. The coexistence of these two local minima persists until $\phi \approx 0.6$, where the disordered phase becomes unstable, nicely matching the experimentally observed coexistence range.

To investigate the nucleation process associated with the first-order phase transition, we examine the evolution of ordered nuclei or clusters in the system with different tapping intensities. The criteria for two particles to belong to the same cluster involve evaluating their distance and orientation. Specifically, for two spherocylinders $i$ and $j$, it is required that their surface distance $d_{ij} < 0.5D$, and their orientation vectors satisfy $|\mathbf{u}_i \cdot \mathbf{u}_j| > 0.995$ [43]. In the inset of Fig. 4(a), we show the size distribution of clusters $P(n)$ at different $\chi$. If the formation of the nuclei is thermally driven based on the Edwards statistics, $P(n)$ at different $\chi$ should be determined by $P(n) = P_0 \exp(-\Delta G(n)/\chi)$, with a normalization factor $P_0$, from which we can infer the free energy $\Delta G(n)$ associated with a cluster of size $n$ [symbols in Fig. 4(a)]. According to the classical nucleation theory (CNT) [44], this free energy $\Delta G(n)$ should consist of at least a bulk and a surface tension term:

$$\Delta G = An(\chi - \chi_0) + Bn^{2/3}. \tag{9}$$

By employing a single set of fitting parameters $A$ and $B$, $\Delta G$ of systems with different $\chi$ can be reasonably fitted [curves in Fig. 4(a)]. Furthermore, the bulk term is given by $An(\chi - \chi_0) = n\Delta G_b$, where $\Delta G_b$ represents the change in free energy per particle at a super-cooling of $(\chi - \chi_0)$. As shown in Fig. 4(c), $\Delta G_b$ decreases as compactivity $\chi$ decreases and becomes zero around $\phi \approx 0.55$. Alternatively, $\Delta G_b$ can also be determined from the free energy $f$, calculated as the difference between the two minimum values corresponding to the disordered state and the ordered nematic phase. It turns out that $\Delta G_b$ obtained from the two approaches are in nice agreement, indicating the validity of CNT even in the athermal granular systems.

In summary, using x-ray tomography, we carry out an experimental study of the nematic phase transition in tapped spherocylinder systems. We first verify the applicability of Edwards ensemble framework to our spherocylindrical system. Subsequently, within this framework, we successfully characterized the first-order nematic phase transition and the associated nucleation process in the system. The experimental results are consistent with the prediction of an empirical free energy that we have introduced. This empirical free energy incorporates terms that account for gravity coupling and the mechanical stability requirements, which are unique to granular materials. Our findings suggest that the phase transition concepts developed in thermal systems can also be generalized to granular materials after appropriate modifications, despite their athermal nature. These results significantly contribute to our understanding of the fundamental principles governing phase transitions in granular systems.

The work is supported by the National Natural Science Foundation of China (No. 12274292 and 11974240), and the Science and Technology Innovation Foundation of Shanghai Jiao Tong University (No. 21X010200829).

Corresponding author

*cjxia@phy.ecnu.edu.cn

†yujiewang@sjtu.edu.cn

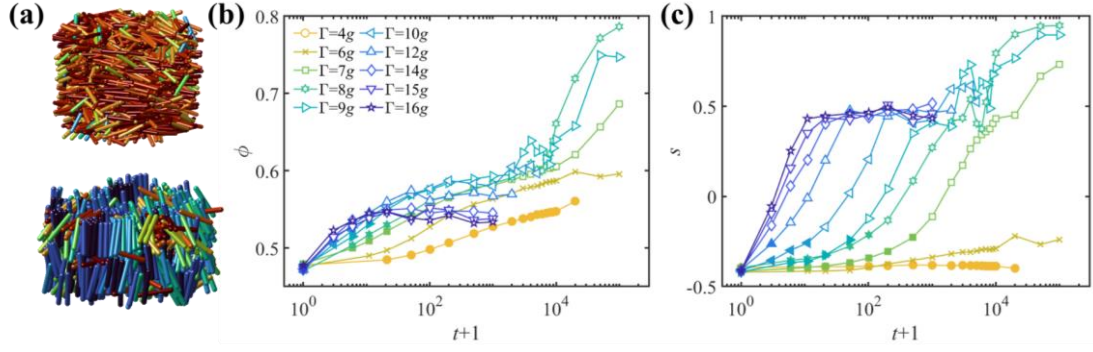

FIG. 1. (a) Reconstructed packing structures in the disordered state (upper, $\Gamma = 8g$, $t=0$) and the nematic phase (lower, $\Gamma = 14g$, $t=10^2$). (b) Packing fraction $\phi$ and (c) nematic order parameter $s$ as a function of the tapping number $t$ at different tapping intensities $\Gamma$. In (b) and (c), the solid symbols represent packings in the disordered state with $s < -0.2$.

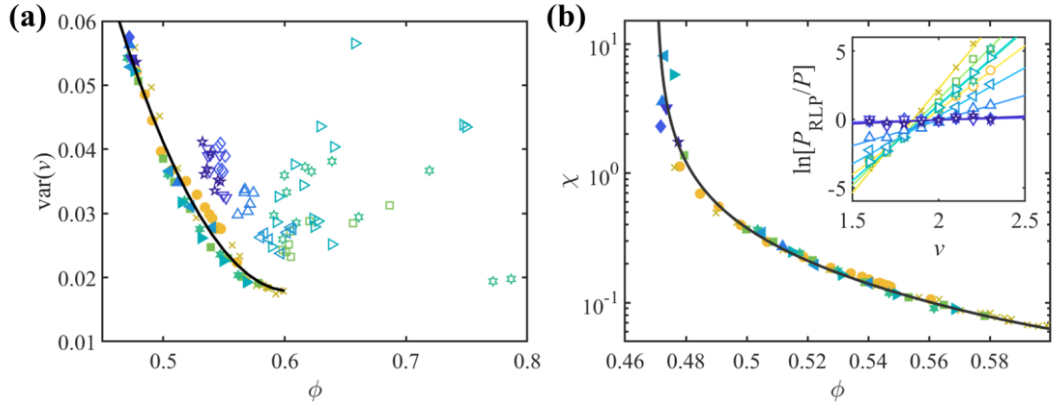

FIG. 2. (a) Voronoi cell volume fluctuation var($v$) as a function of $\phi$ (symbols). The solid symbols represent data for packing in the disordered state. The curve is a cubic polynomial fit for the disordered branch. (b) The compactivity $\chi$ of the disordered packings as a function of $\phi$ calculated via the histogram overlapping method (symbols) and the fluctuation method (solid curve). Inset: the logarithm of the ratios between the volume distributions $P(v)$ of tapped packings under different intensities and the corresponding RLP states.

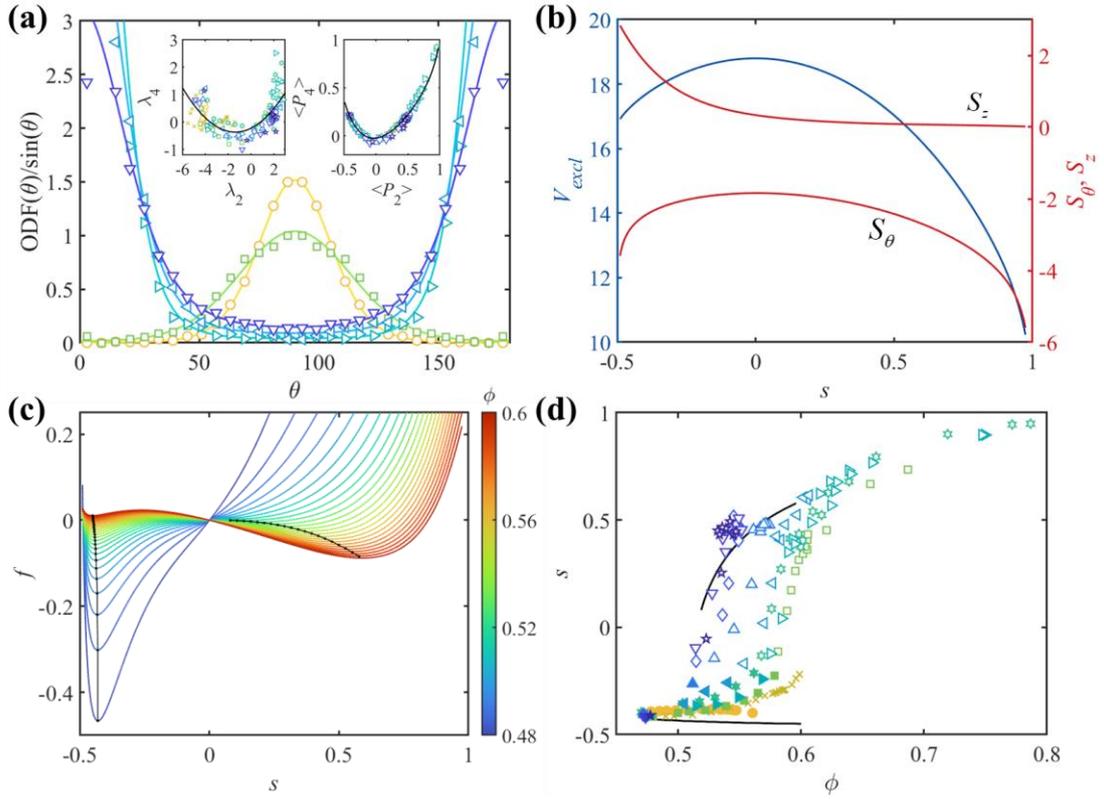

FIG. 3. (a) ODFs for systems in their respective steady states at different $\Gamma$. The left inset shows the relationship between experimental $\lambda_2$ and $\lambda_4$ (symbols), along with the corresponding empirical fitting formula $\lambda_4 = f(\lambda_2)$ (solid curve). The symbols in the right inset are pairs of experimentally obtained $(\langle P_2 \rangle, \langle P_4 \rangle)$ and the solid curve is result derived from the empirical formula $\lambda_4 = f(\lambda_2)$. (b) Average excluded volume $\langle V_{excl} \rangle$ (bule curve, left axis), orientational entropy $S_\theta$ (lower red curve, right axis), and the entropy of contact part $S_z$ (upper red curve, right axis) as a function of $\phi$. (c) Free energy $f$ as a function of $s$ at different $\phi$. The black dots and curves represent the two local minima corresponding to the disordered and nematic phases, respectively. (d) Nematic order parameter $s$ for packings at different $\phi$ (symbols). Data for packings in the disordered state are marked with solid symbols. The solid curves are results derived from the free energy.

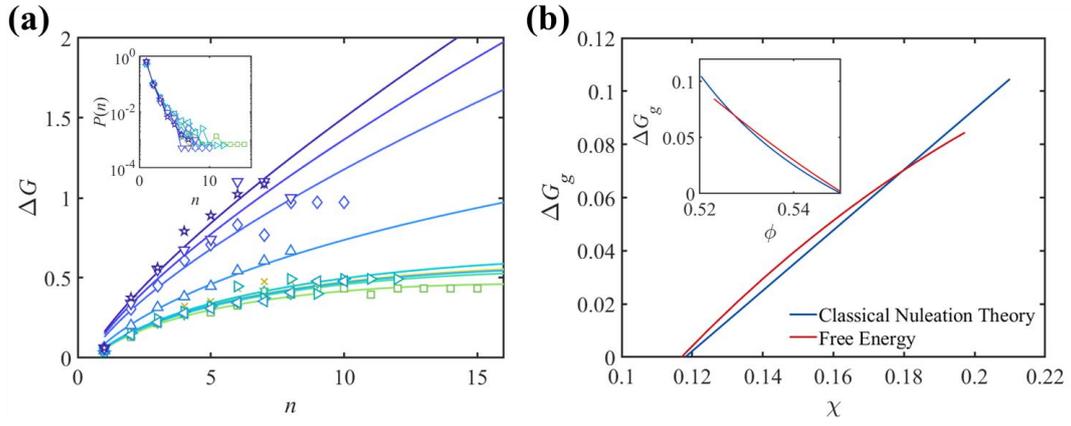

FIG. 4. (a) Free energy $\Delta G$ of forming a cluster with size $n$ calculated through its size distributions (symbols) for packings in the steady state of the nematic phase at different $\Gamma$. The solid curves are fitting results of CNT. Inset: size distributions of clusters $P(n)$ at different $\Gamma$. (b) The change in free energy per particle $\Delta G_b$ as a function of the compactivity $\chi$ calculated via CNT (blue curve) and the modified free energy (red curve). Inset: $\Delta G_b$ as a function of $\phi$ obtained from the two protocols.

# Supplemental Materials for Experimental Study of the Nematic Transition in Granular Spherocylinder Packings under Tapping

Haitao Yu,[1] Zhikun Zeng,[1] Ye Yuan,[1] Shuyang Zhang,[1] Chengjie Xia,[4,*] and Yujie Wang[1,2,3,†]

[1]School of Physics and Astronomy, Shanghai Jiao Tong University, 800 Dong Chuan Road, Shanghai 200240, China

[2]State Key Laboratory of Geohazard Prevention and Geoenvironment Protection, Chengdu University of Technology, Chengdu 610059, China

[3]Department of Physics, College of Mathematics and Physics, Chengdu University of Technology, Chengdu 610059, China

[4]School of Physics and Electronic Science, East China Normal University, Shanghai 200241, China

Corresponding author

*cjxia@phy.ecnu.edu.cn

†yujiewang@sjtu.edu.cn

## 1. Average contact number

We follow previous studies to determine the particle contacts by employing a complementary error function fitting [42]. Figure S1(a) shows that the average contact number $z$ for packings in the disordered state (solid symbols) displays a clear linear relationship with the packing fraction $\phi$, whereas $z$ decreases for those in the nematic phase with higher order parameter $s$. Consequently, we develop an empirical expression to model the relationships between z and both $\phi$ and $s$ for all packings, as shown in Fig. S1(b):

$$z(\phi, s) = 22.12\phi - 0.2354\exp(3.32s) - 3.873 . \tag{S1}$$

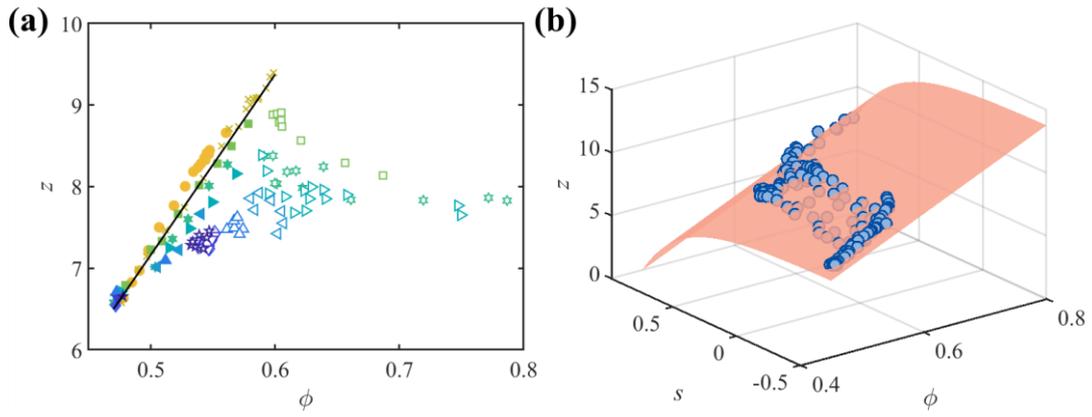

FIG. S1. (a) Average contact number $z$ as a function of packing fraction $\phi$. The solid line is the linear fitting of data for packings in the disordered state (solid symbols). (b) Average contact number $z$ for packings at different $\phi$ and $s$ (symbols). The red surface is the empirical fitting of $z(\phi, s)$.